\documentclass[letters,fleqn,usenatbib]{mnras}

\usepackage{newtxtext,newtxmath}

\usepackage[T1]{fontenc}

\DeclareRobustCommand{\VAN}[3]{#2}
\let\VANthebibliography\thebibliography
\def\thebibliography{\DeclareRobustCommand{\VAN}[3]{##3}\VANthebibliography}

\usepackage{graphicx}	
\usepackage{amsmath}	
\usepackage{wasysym}
\usepackage{newtxtext,newtxmath}
\usepackage{bm}

\newcommand{\pyaneti}{\texttt{pyaneti}}

\newcommand{\lbe}{$\lambda_{\rm e}$}
\newcommand{\lbp}{$\lambda_{\rm p}$}
\newcommand{\pgp}{$P_{\rm GP}$}

\newcommand{\lrv}{$\ln \mathcal{L}_\mathrm{rv}^\prime$}

\newcommand{\mpgicrv}{$\mathrm{MGIC}_{\rm rv}$}
\newcommand{\ktot}{$\mathcal{K}_{\rm rv}$}
\newcommand{\ksmooth}{$\mathcal{K}_{\rm s}$}

\newcommand{\monthyear}{\@monthname~\number\year}

\newcommand{\aione}{$\mathcal{A}_{1}$}
\newcommand{\aitwo}{$\mathcal{A}_{2}$}

\title[A Model Selection Criterion for Multi-GPs]{A Model Selection Criterion for Multidimensional Gaussian Processes: Application to Radial Velocities}

\author[Barragán]{
Oscar~Barragán$^{1,2}$\thanks{\href{mailto:oscar.barragan@warwick.ac.uk}{oscar.barragan@warwick.ac.uk}}
\\
$^{1}$ Department of Physics, University of Warwick, Coventry, CV4 7AL, UK \\
$^{2}$ Sub-department of Astrophysics, Department of Physics, University of Oxford, Oxford, OX1 3RH, UK \\
}

\date{Accepted XXX. Received YYY; in original form ZZZ}

\pubyear{\the\year{}}

\begin{document}
\label{firstpage}
\pagerange{\pageref{firstpage}--\pageref{lastpage}}
\maketitle

\begin{abstract}
Multidimensional Gaussian Process (multi-GP) regression is widely used to disentangle stellar and planetary signals in radial velocities (RVs) by jointly modelling ancillary activity indicators. However, identifying the combination of indicators that best constrains the stellar signal in the RVs is non-trivial, as classical model comparison methods are not directly applicable when multi-GPs involve different  time series combinations.
In this work, we present an information criterion to compare multi-GP models based on their ability to explain the RV component, \mpgicrv. 
This metric combines the conditional RV likelihood with an effective parameter count that accounts for the regularisation imposed by the multi-GP model on the RV component.
We demonstrate that \mpgicrv\ provides a quantitative and robust framework for multi-GP model comparison, identifying the activity indicators that most effectively constrain the RV signal. 
Although developed in the context of RV analysis, the proposed criterion is general and applicable to multi-GP problems in which the inference focuses on a specific observable.
\end{abstract}




\section{Introduction}  

One of the central challenges in the detection of exoplanets using the radial velocity (RV) technique is the presence of stellar signals in the spectroscopic data. 
In particular, RV variations induced by stellar activity can mimic or obscure the Doppler signals caused by orbiting planets, limiting the detectability and the reliability of planetary detections \citep[e.g.,][]{Queloz2001,Rajpaul2016}. 
These activity-induced signals arise from phenomena such as spots, plages, and magnetic cycles, and typically evolve on timescales comparable to stellar rotation and active-region lifetimes \citep[see e.g.,][]{Dumusque2014,Rajpaul2015}. 
The most popular methods to deal with this are based on modelling the activity-induced variability alongside potential planetary signals in the RV timeseries \citep[e.g.,][]{Haywood2014,Rajpaul2015}. In this context, Gaussian Processes (GPs) are among the preferred approaches.

Gaussian Processes are well suited to modelling stellar signals because {they provide a flexible representation of time correlations} \citep[e.g.,][]{Haywood2014,Aigrain2022}. 
However, this property can become a drawback if the GP is not sufficiently constrained, potentially leading to overfitting or the absorption of planetary signals into the activity model. 
To avoid this issue, \citet[][]{Rajpaul2015} proposed a multidimensional GP (multi-GP) regression approach that exploits the relation between RVs and ancillary activity indicators \citep[based on][]{Aigrain2012} to constrain the stellar component in the RVs. 
Multi-GP regressions have since been shown to be effective in disentangling planetary and stellar signals in RV data spanning a wide range of stellar activity levels \citep[e.g.,][]{Barragan2023,Barragan2026,Luque2023}.

Recent studies have shown that one of the advantages of multi-GP regressions, in comparison to pure onedimensional (1D) ones, is that they can lead to more precise planetary parameter estimates \citep[e.g.,][]{pyaneti2,Barragan2023}. 
While these results demonstrate the advantages of informing the RV time series with contemporaneous activity indicators, it is still not clear how to choose the optimal combination of time series that best constrain the RV stellar activity signal. This could be done by quantitative model comparison. However, in the context of multi-GP regression this is a non-trivial problem. 
Classical model comparison criteria, such as the Akaike Information Criterion \citep[AIC;][]{Akaike1974} or Bayesian evidence, are designed to compare models applied to the same data set. 
As a result, they cannot be directly applied to multi-GP models that incorporate different time series combinations \citep[see][]{Ahrer2021,Rajpaul2021,Barragan2023}. 
{  
Several attempts have been made in the literature to perform such comparisons \citep[e.g.,][]{Rajpaul2021,Zhao2022}, but none has proved fully satisfactory. More recently, \citet{Hara2025} demonstrated that cross-validation can be used for model selection in multi-GP analyses. As cross-validation is the standard approach for comparing GP models through their predictive performance on held-out data \citep[][]{Rasmussen2006}, this result provides a promising avenue for multi-GP model comparison. Furthermore, cross-validation and AIC have been shown to be asymptotically equivalent for model selection \citep[e.g.,][]{Stone1977}. 
Motivated by this connection, we introduce an AIC-like information criterion that exploits the joint multivariate Gaussian structure of multi-GPs to compare models according to their ability to explain the RV component.
}

\section{An Information Criterion for multi-GPs}
\label{sec:aicequations}

In this section, we introduce a model comparison framework tailored to multi-GP regressions applied to RVs and stellar activity indicators. 
We focus on the ability of each model to explain the RV observations. 
Our goal is to compare models that may use different combinations of activity time series, while ensuring that the comparison is performed on a common basis.
We adopt an heuristic approach inspired by the AIC \citep[][]{Akaike1974}. 
Conceptually, the AIC consists of two parts: a likelihood term measuring how well the model explains the observations, and a penalty term quantifying the number of parameters used by the model.
In the next subsections we describe how to compute these two terms to perform multi-GP comparison.

\subsection{Conditional likelihood on the RVs}

The central strategy of multi-GP regression is to reformulate the original multidimensional inference as a single, large-scale 1D GP regression. This is done by assembling a \emph{big} residual vector, $\bm{r}_{\rm big}$, and a corresponding covariance matrix, $\mathbf{K}_{\rm big}$ to build a likelihood \citep[see,][]{Rajpaul2015,pyaneti2}.
The residual vector $\bm{r}_{\rm big}$ is constructed by concatenating the residual vectors for each dimension $l$: $\bm{r}_l = \bm{y}_l - \bm{\mu}_l$, where the $\bm{y}_l$ and $\bm{\mu}_l$ vectors denote the observation and mean vectors for each dimension $l$, respectively. The associated covariance matrix $\mathbf{K}_{\rm big}$ is assembled from sub-matrices $\mathbf{K}_{l,m}$, which capture all  covariances between time series $l$ and $m$. 
We can partition the multi-GP framework into two blocks: one corresponding to the RVs and the other to the full set of activity indicators. We will use the subscript ``rv'' to refer to the RV dimension, and ``ai'' to denote the combined activity indicator dimensions. The observed data vector can then be written as
\begin{equation}
    \bm{y}_{\rm big} =
    \begin{pmatrix}
    \bm{y}_{\rm rv} \\
    \bm{y}_{\rm ai} \\
    \end{pmatrix}
    ,
    \label{eq:bigr}
\end{equation}
\noindent
the mean vector
\begin{equation}
    \bm{\mu}_{\rm big} =
    \begin{pmatrix}
    \bm{\mu}_{\rm rv} \\
    \bm{\mu}_{\rm ai} \\
    \end{pmatrix}
    ,
    \label{eq:bigmu}
\end{equation}
\noindent
and the covariance matrix
\begin{equation}
    \bm{K}_{\rm big} =
    \begin{pmatrix}
    \bm{K}_{\rm rv,rv} & \bm{K}_{\rm rv,ai}  \\
    \bm{K}_{\rm ai,rv}  & \bm{K}_{\rm ai,ai} \\
    \end{pmatrix}
    .
    \label{eq:bigk}
\end{equation}
{  The $\bm{K}_{l,m}$ matrices are constructed as
\begin{equation}
    \bm{K}_{l,m} = \bm{k}_{l,m} + \bm{R}_{l,m},
\end{equation}
where $\bm{k}_{l,m}$ matrices are obtained by evaluating the kernel function at all pairs of observation times in dimensions $l$ and $m$, and $\bm{R}_{l,m}$ represents the observational noise covariance, with $\bm{R}_{l,m} = \bm{0}$ for $l \neq m$ and diagonal blocks $\bm{R}_{l,l}$ containing the measurement uncertainties and jitter terms.}

For our model comparison, we are interested in computing the conditional probability of observing $\bm{y}_{\rm rv}$ given the observations $\bm{y}_{\rm ai}$, written as $P(\bm{y}_{\rm rv} \mid \bm{y}_{\rm ai})$. For a multivariate Gaussian distribution, this is computed as \citep[see][]{anderson1984,styan1985}
\begin{equation}
P(\bm{y}_{\rm rv}\mid \bm{y}_{\rm ai}) = \mathcal{N}\big(\bm{\mu}_{\rm rv}^\prime, \bm{K}_{\rm rv,rv}^{\prime}\big),
\label{eq:conditional}
\end{equation}
where
\begin{equation}
\bm{\mu}_{\rm rv}^\prime = \bm{\mu}_{\rm rv} + \bm{K}_{\rm rv,ai} \bm{K}_{\rm ai,ai}^{-1} (\bm{y}_{\rm ai} - \bm{\mu}_{\rm ai}),
\label{eq:rprime}
\end{equation}
is the conditional mean vector for the RVs, and
\begin{equation}
\bm{K}_{\rm rv,rv}^\prime = \bm{K}_{\rm rv,rv} - \bm{K}_{\rm rv,ai} \bm{K}_{\rm ai,ai}^{-1} \bm{K}_{\rm ai,rv},
\label{eq:schur}
\end{equation}
is the conditional covariance of the RVs, given by the Schur complement of the $\bm{K}_{\rm ai,ai}$ sub-matrix \citep[see][]{styan1985,Lu2002}.

We can see eqs.~\eqref{eq:rprime} and~\eqref{eq:schur} as corrections to the unconditioned mean and covariance of the RVs.
The terms involving $\bm{K}_{\rm rv,ai}\bm{K}_{\rm ai,ai}^{-1}$ account for the part of the variability in the RVs that can be \emph{explained} by knowing the activity indicators.
Equation~\eqref{eq:rprime} accounts for how well the joint model explains $\bm{y}_{\rm ai}$, and eq.~\eqref{eq:schur} quantifies the extent to which the variability in $\bm{y}_{\rm rv}$ is reduced by knowing $\bm{y}_{\rm ai}$.
If there is no connection between RVs and activity indicators {  (i.e., $\bm{K}_{\rm rv,ai} = \bm{0}$)} we recover the RVs unconditioned mean vector, $\bm{\mu}_{\rm rv}^\prime = \bm{\mu}_{\rm rv}$, and covariance matrix, $\bm{K}_{\rm rv,rv}^\prime = \bm{K}_{\rm rv,rv}$.

We can then compute the conditional likelihood for the RV observations, accounting for their joint distribution with the activity indicators, as
\begin{equation}
\ln \mathcal{L}_\mathrm{rv}^\prime =
- \frac{1}{2} \left(
N_{\rm rv} \ln 2\pi + \ln \lvert \bm{K}_{\rm rv,rv}^\prime \rvert + {\mathbf{r}_{\mathrm{rv}}^\prime}^{\intercal} {\bm{K}_{\rm rv,rv}^\prime}^{-1} \mathbf{r}^{\prime}_{\mathrm{rv}}
\right),
\label{eq:loglikelihood}
\end{equation}
\noindent
where 
\begin{equation}
\bm{r}^{\prime}_{\rm rv} = \bm{y}_{\rm rv} - \bm{\mu}_{\rm rv}^\prime, 
\end{equation}
\noindent
and $N_{\rm rv}$ is the number of RV observations. Note how eq.~\eqref{eq:loglikelihood} does not depend on the  number of dimensions or on the number of activity indicator observations. 
This likelihood computes how well the RV part of the model explains $\bm{y}_{\rm rv}$, incorporating the information provided by the activity indicators.
However, this alone is insufficient for model comparison, models that incorporate additional activity indicators or more flexible covariance structures can achieve higher likelihood values simply through increased functional flexibility.
We still need a penalty that balances goodness of fit against model complexity.

\subsection{Total number of parameters affecting the RV dimension}

In classical parametric models, the complexity penalty in the AIC is proportional to the number of free parameters. 
However, in multi-GP models, complexity arises from two sources. First, the explicit number of model parameters $\mathcal{K}_{\rm p}$ entering in the likelihood computation. This includes the mean function parameters (e.g., Doppler semi-amplitudes, offsets, etc.), kernel hyperparameters, jitter terms. 
The second source of complexity comes from the flexibility of the GP component itself.

To estimate the number of parameters due to the flexibility of a multi-GP, we will use the property of GPs being linear {  smoothers} \citep[see e.g.,][]{Rasmussen2006}.
This means that the mean values of the big predictive distribution at the observing times, $\widehat{\bm{y}}_{\rm big}$, are computed as a linear transformation of the observations as
\begin{equation}
\widehat{\bm{y}}_{\rm big}
=
\bm{\mu}_{\rm big}
+
\mathbf{S}_{\rm big}
\left(\bm{y}_{\rm big} - \bm{\mu}_{\rm big}\right),
\end{equation}
where
\begin{equation}
\mathbf{S}_{\rm big}
=
\bm{k}_{\rm big} \bm{K}_{\rm big}^{-1}  
,
\label{eq:smoother_full}
\end{equation}
is the so-called \emph{smoother} (or \emph{hat}) matrix,
\begin{equation}
    \bm{k}_{\rm big} =
    \begin{pmatrix}
    \bm{k}_{\rm rv,rv} & \bm{k}_{\rm rv,ai}  \\
    \bm{k}_{\rm ai,rv}  & \bm{k}_{\rm ai,ai} \\
    \end{pmatrix}
    ,
    \label{eq:bigkcov}
\end{equation}
{  is the kernel-only covariance matrix between all dimensions (i.e., without the noise component), and $\bm{K}_{\rm big}$ is given in eq.~\eqref{eq:bigk}.}
The smoother matrix characterises the linear mapping between observational residuals and predicted values, quantifying the sensitivity of the model predictions to the data. In this sense, $\mathbf{S}_{\rm big}$ encodes the flexibility of the latent GP component. 
{  Larger diagonal elements would correspond to stronger responsiveness of the predictive values to individual observations, while smaller values reflect stronger regularisation. For this reason, the degrees of freedom of linear smoothers (like GPs) are obtained from the trace of the smoother matrix \citep[for further details see e.g.,][]{Ye1998,hastie1990,Rasmussen2006}.}
We would expect that if a given set of activity indicators truly explains the stellar signal, the RV component of the model should require fewer effective degrees-of-freedom (i.e. less flexibility) to explain $\bm{y}_{\rm rv}$. 
Since our objective is to compare multi-GP models based on their ability to explain RVs, we will use the smoother matrix $\mathbf{S}_{\rm rv}$, that relates $\bm{y}_{\rm rv}$ with their corresponding predictive values
\begin{equation}
\widehat{\bm{y}}_{\rm rv}
=
\bm{\mu}_{\rm rv}
+
\mathbf{S}_{\rm rv}
\left(\bm{y}_{\rm rv} - \bm{\mu}_{\rm rv}\right),
\end{equation}
where (see Appendix~\ref{ap:hrv})
\begin{equation}
    \mathbf{S}_{\rm rv} =
    \bm{k}_{\rm rv,rv} \bm{K}_{\rm rv,rv}^{\prime \, -1} - \bm{k}_{\rm rv,ai}\bm{K}_{\rm ai,ai}^{-1}\bm{K}_{\rm ai,rv}\bm{K}_{\rm rv,rv}^{\prime \, -1},
    \label{eq:hatrvrv}
\end{equation}
\noindent
and the smoothing-based degrees-of-freedom for the RVs is then 
\begin{equation}
\mathcal{K}_{\rm s} = \mathrm{tr}\left(\mathbf{S}_{\rm rv}\right).
\label{eq:df_smooth}
\end{equation}

Equation~\eqref{eq:hatrvrv} shows how conditioning on the activity indicators reduces the effective covariance of the RV component. In the limiting case where the activity indicators fully capture the stellar signal present in the RVs, the diagonal elements of $\mathbf{S}_{\rm rv}$ approach to zero, and $\mathrm{tr}(\mathbf{S}_{\rm rv})$ tends to zero. Thus, $\mathbf{S}_{\rm rv}$ quantifies the extent to which the activity indicators regularise the RV activity signal and can be seen as a generalisation of the number of parameters to account for the model flexibility in the RV component.
If we combine eq.~\eqref{eq:df_smooth} with the number of model parameters, $ \mathcal{K}_{\rm p}$, we can define the total number of effective parameters influencing the RV dimension
\begin{equation}
\mathcal{K}_{\rm  rv}=
\mathcal{K}_{\rm  p}
+
\mathcal{K}_{\rm  s}.
\label{eq:keff_final}
\end{equation}

\subsection{An AIC-like information-criterion for multi-GPs}
\label{sec:aicformultis}

Having defined the conditional RV likelihood \lrv\ { (computed from the best fit parameters)}, and the effective parameter count for the RV dimension \ktot, we can now construct a multi-GP information criterion applied to the RV dimension as
\begin{equation}
\mathrm{MGIC}_{\rm rv}
=
-2 \ln \mathcal{L}_{\rm rv}^{\prime}
+
2 \mathcal{K}_{\rm rv}.
\label{eq:mpgicrv}
\end{equation}

We emphasise that this construction is heuristic. 
While the conditional likelihood is exact under the assumed Gaussian model, the number of parameters in a multi-GP setting is not uniquely defined. 
The smoothing-based degrees-of-freedom term, \ksmooth, provides a principled approximation of model flexibility, extending established results for linear smoothers to multi-GPs. 
We will interpret the unit differences in \mpgicrv\ as for the AIC. Differences less than 2 units are negligible, differences around 4 units offer some evidence in favour of the model with the lower value, and differences greater than 10 units provide strong evidence \citep{Burnham2002}. 
{  These thresholds are compatible with the values we obtain in the criterion test in Sect.~\ref{sec:synthetic}.}

\section{Testing $\mathrm{MGIC}_{\rm rv}$}
\label{sec:testmetric}

\subsection{Test on synthetic data}
\label{sec:synthetic}

{ 
To evaluate the performance of \mpgicrv\ introduced in Eq.~\eqref{eq:mpgicrv}, we designed synthetic data sets based on a multi-GP model. We follow the definitions of photometric and RV-like activity indicators introduced by \cite{Barragan2023}. Specifically, we generated two synthetic time series: $\mathcal{A}_{1}$, behaving as an RV-like signal, and $\mathcal{A}_{2}$, behaving as a photometry-like activity indicator. These time series are defined as
\begin{equation}
\begin{aligned}
\mathcal{A}_{1}  &= A_{1} G(t) + B_{1} \dot{G}(t) + \varepsilon_1(t), \\
\mathcal{A}_{2} &= A_{2} G(t) + \varepsilon_2(t),
\end{aligned}
\end{equation}
where $\varepsilon_i(t)$ is a white noise term, the parameters $A_{i}$ and $B_{i}$ serve as free coefficients that connect each time-series to the latent process. Here, $G(t)$ is a latent function sampled using a quasi-periodic (QP) kernel defined as
\begin{equation}
   \gamma_{{\rm QP}} (t_i, t_j) = \exp 
    \left[
    - \frac{\sin^2[\pi(t_i - t_j)/P_{\rm GP}]}{2 \lambda_{\rm P}^2}
    - \frac{(t_i - t_j)^2}{2 \lambda_{\rm e}^2}
    \right],
    \label{eq:gamma}
\end{equation}
where $P_{\rm GP}$ is the characteristic GP period; $\lambda_{\rm P}$ is the inverse of the harmonic complexity; and $\lambda_{\rm e}$ is the evolution timescale.

We generated synthetic time series using the \texttt{citlalatonac} code \citep{pyaneti2}. For the QP kernel, we adopted \pgp~$=10$\,d, \lbp$=0.7$, and \lbe$=25$\,d. A total of 50 observations were simulated at random epochs within a 50\,d window.
We varied the coefficients $A_1$, $B_1$, and $A_2$ between 0 and 5 in steps of 1, resulting in 216 different time series combinations. All units are arbitrary. We added Gaussian white noise with standard deviation of 1 to all time series. These configurations include cases in which the time series are dominated by white noise, as well as scenarios where $\mathcal{A}_{1}$ is dominated by either the $G(t)$ or $\dot{G}(t)$ component.

Our goal is to assess whether \mpgicrv\ can correctly identify the multi-GP model that best describes the $\mathcal{A}_{1}$ time series. For each of the 216 synthetic data sets, we therefore tested five model configurations. 
Model~1 consists of a one-dimensional GP regression on $\mathcal{A}_{1}$ only, assuming $\mathcal{A}_{1} = A_{1}G(t)$. 
Model~2 corresponds to describing $\mathcal{A}_{1}$ as $A_{1}G(t) + B_{1}\dot{G}(t)$ and $\mathcal{A}_{2}$ as $A_{2}G(t)$ (the true underlying model).
Model~3 describes both $\mathcal{A}_{1}$ and $\mathcal{A}_{2}$ as combinations of $G(t)$ and $\dot{G}(t)$, making it more complex than the true model. Model~4 assumes that both $\mathcal{A}_{1}$ and $\mathcal{A}_{2}$ depend only on $G(t)$, neglecting the contribution from $\dot{G}(t)$. Finally, Model~5 corresponds to describing $\mathcal{A}_{1}$ as $A_{1}S(t) + B_{1}\dot{S}(t)$ and $\mathcal{A}_{2}$ as $A_{2}S(t)$, where $S(t)$ is generated using a squared-exponential (SE) kernel \citep[as defined in][]{pyaneti2}.
All models are summarised in Table~\ref{tab:synthetic_models}. Together, these configurations span true, oversimplified, over-complex, and incorrect models.

\begin{table}
\centering
\caption{Synthetic multi-GP models tested in this work.}
\label{tab:synthetic_models}
\begin{tabular}{cl}
\hline
Model & Description \\
\hline
1 & $\mathcal{A}_{1} = A_{1}G(t)$,\,  \\
2 & $\mathcal{A}_{1} = A_{1}G(t) + B_{1}\dot{G}(t)$,\,  $\mathcal{A}_{2} = A_{2}G(t)$ \\
3 & $\mathcal{A}_{1} = A_{1}G(t) + B_{1}\dot{G}(t)$,\, $\mathcal{A}_{2} = A_{2}G(t) + B_{2}\dot{G}(t)$ \\
4 & $\mathcal{A}_{1} = A_{1}G(t)$,\, $\mathcal{A}_{2} = A_{2}G(t)$ \\
5 & $\mathcal{A}_{1} = A_{1}S(t) + B_{1} \dot{S}(t)$,\, $\mathcal{A}_{2} = A_{2} S(t)$ \\
\hline
\end{tabular}
\end{table}

We ran a series of multi-GP regressions using \pyaneti\ \citep[][]{pyaneti,pyaneti2}.
We used 250 walkers in the ensemble Markov chain Monte Carlo (MCMC) sampler \citep{emcee}. Posterior distributions were constructed from the final 5000 iterations of the converged chains. Thinning the chains by a factor of 10 resulted in 500 samples per walker, corresponding to 125,000 posterior samples per parameter.
We sampled for the QP kernel hyperparameters, the corresponding multi-GP amplitudes, and an offset and jitter term for each time series. We adopted broad uniform priors for all the parameters.

\begin{figure*}
\centering
\includegraphics[width=\linewidth]{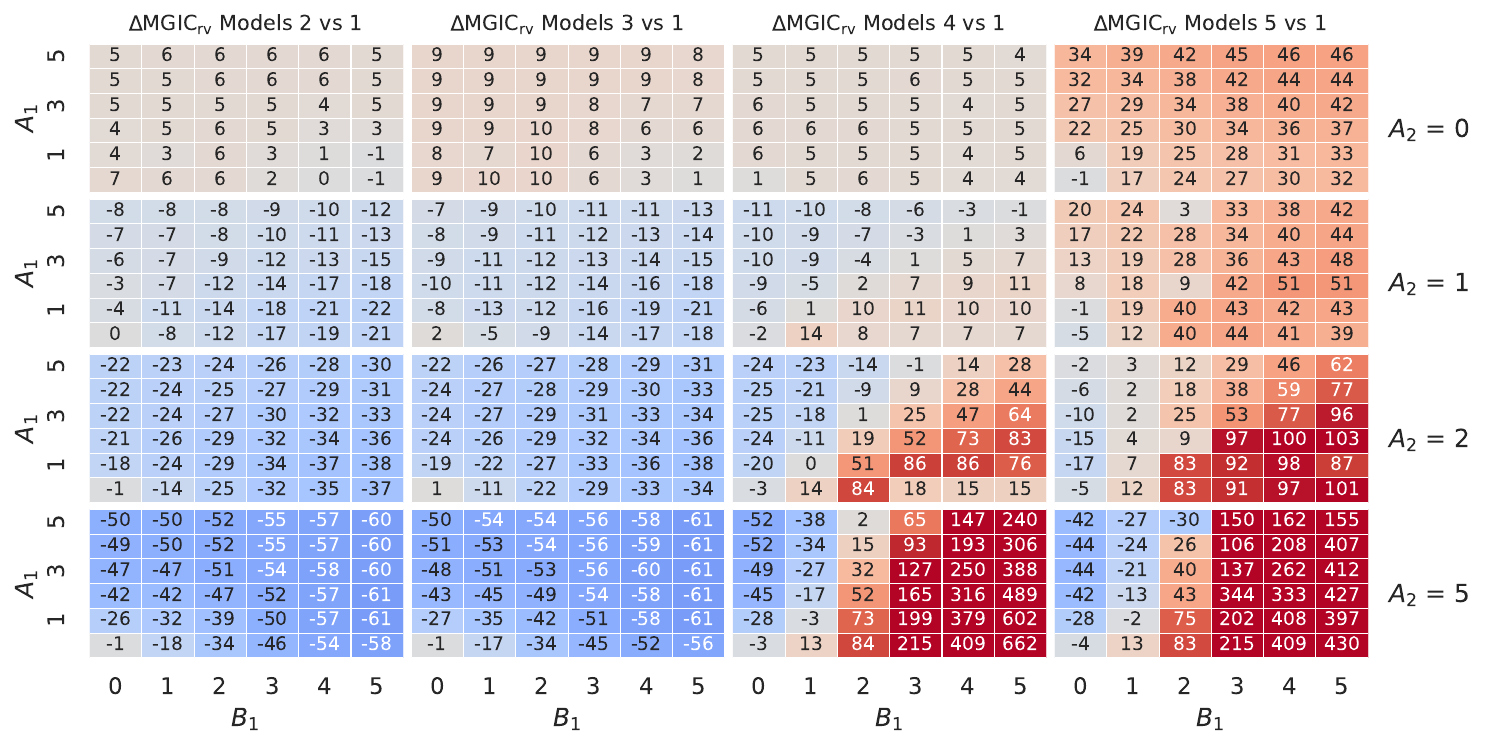}\\
\caption{Heat maps of $\Delta$\mpgicrv\ quantities as functions of $A_1$ and $B_1$ for varying $A_2$. Each row corresponds to a fixed $A_2$, and each column to a different $\Delta$\mpgicrv\ taking Model 1 as reference. Colours represent normalised delta magnitudes, with annotations showing the exact  $\Delta$\mpgicrv\ values.}
\label{fig:mgic_all_tests}
\end{figure*}

Figure~\ref{fig:mgic_all_tests} shows the $\Delta$\mpgicrv\ values obtained for the different model comparisons, using Model~1 as the reference (See Appendix~\ref{sec:model2comparison} for a brief discussion using Model~2 as reference). We choose Model~1 because it represents the standard 1D GP regression commonly used in RV analyses. This comparison therefore illustrates how, and under which conditions, multi-GP models improve upon a purely 1D treatment of the RV time series. It also shows how \mpgicrv\ responds to the inclusion of ancillary indicators.

The first notable result is that \mpgicrv\ strongly penalises Model~5, despite it having the same functional form as the true model. Model~5 differs only in its covariance structure by assuming a SE kernel, while the data were generated with a QP kernel. This is reflected in the large $\Delta$\mpgicrv\ between Model~5 and Model~1 (column~4 of Fig.~\ref{fig:mgic_all_tests}). Model~5 is preferred over Model~1 only in the limiting cases where the activity indicator has a strong signal ($A_2=5$) and the derivative contribution to the RV-like time series is negligible ($B_1=0$). Even in these cases, however, Model~5 remains disfavoured relative to Models~2 and~3 (see Fig.~\ref{fig:mgic_all_tests} and Appendix~\ref{sec:model2comparison}). This shows that \mpgicrv\ is sensitive not only to the functional structure of the multi-GP model (see below), but also to the choice of kernel. This suggest that the criterion can therefore also help identify the covariance structure that best describes the data. From now on, we will discuss only the models created using the correct kernel function.

Another important result is that when the activity indicator is dominated by white noise (i.e. $A_2 = 0$, first row of Fig.~\ref{fig:mgic_all_tests}), \mpgicrv\ disfavours the inclusion of \aitwo. In fact, all multi-GP models are slightly penalised relative to the 1D case due to their increased complexity without a corresponding improvement in the \aione\ modelling. As the signal-to-noise of \aitwo\ increases (larger $A_2$ values, we do not show the results for $A_2$ values of 3 and 4, but they follow a similar trend), then \aitwo\ begins to affect the inference on the \aione\ component, and these effects are captured by \mpgicrv.
A similar behaviour is observed when \aione\ itself becomes noise dominated (i.e. $A_1 = B_1 = 0$). In this case, the inclusion of \aitwo\ does not improve the modelling of \aione, regardless of the quality of the activity indicator. This suggest that informative ancillary time series cannot compensate for the absence of recoverable information in the target dimension itself.
The comparison between Models~1 and~2 (first column of Fig.~\ref{fig:mgic_all_tests}) shows that \mpgicrv\ consistently favours Model~2, which corresponds to the true model by construction. The preference becomes stronger as the signal-to-noise increases in both time series.
Model~3 extends Model~2 by including an additional $\dot{G}(t)$ component in \aitwo, introducing unnecessary complexity. Figure~\ref{fig:mgic_all_tests} shows that this additional flexibility does not significantly improve \mpgicrv, and Model~2 is therefore preferred.
Finally, Model~4 provides the clearest illustration of the diagnostic power of \mpgicrv. This model is correct only when $B_1 = 0$, i.e. when \aione\ does not contain a $\dot{G}(t)$ contribution. In this regime, \mpgicrv\ correctly favours Model~4. However, as the contribution from $\dot{G}(t)$ becomes significant, the model rapidly becomes disfavoured. This demonstrates that \mpgicrv\ is not only sensitive to over-complexity, but can also identify activity indicators and/or models that do not fully capture the underlying process governing the RV-like signal.

We also performed similar simulations including injected planetary signals in the RV-like dimension. In all cases, \mpgicrv\ continued to favour the true underlying model, which also gave the most precise recovery of the Doppler semi-amplitude. This behaviour is consistent with previous qualitative results \citep[see][ and next section]{pyaneti2,Barragan2023}.
This shows how \mpgicrv\ is also sensitive to the mean function values via \lrv.

Note that these tests are intended to illustrate the general behaviour of \mpgicrv\ under different scenarios within a specific synthetic configuration. Although similar trends are expected under comparable conditions, the quantitative values of \mpgicrv\ will depend on the number of observations, the sampling strategy, and the adopted kernel and hyper-parameters.

}

\subsection{Test on real systems}

We have shown how the \mpgicrv\ metric can be used to compare multi-GP models to find the best combination of activity indicators when we know the underlying truth. We then tested the behaviour of \mpgicrv\ in real data sets.
For all cases, we use the same MCMC configuration in \pyaneti\ as in Sect.~\ref{sec:synthetic}.
The full list of systems, multi-GP combinations and \mpgicrv\ metrics are given in Table~\ref{tab:results_real}.

\begin{table}
\begin{center}
\caption{\lrv, $\mathcal{K}_{\rm p}$, \ksmooth, and \mpgicrv\ for multi-GPs for real data sets.\label{tab:results_real}}
\begin{tabular}{lcccc}
\hline\hline
 time-series & \lrv & $\mathcal{K}_{\rm p}$ & \ksmooth\ & \mpgicrv  \\
\hline
\multicolumn{5}{c}{K2-233} \\
\hline
 RV  & 475 & 15 & 59 & -802 \\ 
 RV + FWHM  & 492 & 21 & 55 & -831 \\ 
 RV + FWHM + BIS & 493 & 25 & 46 & -845 \\ 
\hline
\multicolumn{5}{c}{TOI-837} \\
\hline
 RV  & 120 & 7 & 15 & -195 \\ 
 RV + FWHM  &  126 & 11 & 20 & -190 \\ 
 RV + FWHM + BIS & 128 & 15 & 24 & -174 \\ 
\hline
\multicolumn{5}{c}{TOI-451} \\
\hline
 RV  & 358 & 15 & 88 & -509 \\ 
 RV + FWHM  & 379 & 19 & 85 & -552 \\ 
 RV + NGTS &  383 & 19 & 84 & -561 \\ 
\hline
\end{tabular}
\end{center}
\end{table}
 
We first revisit the results of \citet{Barragan2023} for K2-233 using \mpgicrv.
Hereafter, We adopt the terminology, where FWHM denotes the full width at half maximum and BIS the bisector inverse slope. We re-run their 1D (RV), 2D (RV + FWHM), and 3D (RV + FWHM + BIS) multi-GP models following the configuration described in their paper. 
They concluded that the best model is the 3D (RV + FWHM + BIS) multi-GP based on the ability of the model to give precise Doppler semi-amplitudes.
As shown in Table~\ref{tab:results_real}, the model with the lowest \mpgicrv\ is the 3D-GP regression including both FWHM and BIS. While the conditional RV likelihood \lrv\ is similar for the 2D and 3D cases, the improvement in \mpgicrv\ arises from a reduction in \ksmooth, indicating stronger regularisation of the stellar activity signal when BIS is included. In this case, \mpgicrv\ quantitatively agrees with the qualitative assessment of \citet{Barragan2023}, demonstrating that the additional indicators constrain the RV activity component for this particular data set.

We next analyse the TOI-837 data set from \citet{Barragan2024}, who qualitatively argued that the activity indicators do not add information in a multi-GP framework for the corresponding data set. We repeat their 1D (RV), 2D (RV + FWHM), and 3D (RV + FWHM + BIS) analyses under the same setup. We find that the 1D model is favoured by \mpgicrv. Although \lrv\ remains similar across models, the inclusion of activity indicators increases \ksmooth. This suggests that additional flexibility is required from the RV dimension to explain the activity signal in the indicator time series. Suggesting that activity indicators do not add information for the TOI-837 RV data set. Once again, the quantitative behaviour of \mpgicrv\ reproduces the qualitative conclusions of the original study.

Finally, we consider the multi-GP analysis of TOI-451 presented by \citet{Barragan2026}, where contemporaneous Next-Generation Transit Survey \citep[NGTS;][]{wheatley18ngts} photometry was found to constrain the RV signal more effectively than spectroscopic indicators. We re-ran their 1D (RV), 2D (RV + FWHM), and 2D (RV + NGTS) models using their same configuration. As reported in Table~\ref{tab:results_real}, both 2D models improve upon the 1D case, and the RV + NGTS model yields the lowest \mpgicrv, quantitatively confirming that contemporaneous photometry provides a slightly best constraint on the RV activity component for this system.

Across these diverse systems, \mpgicrv\ consistently identifies the activity indicators that most effectively constrain the RV signal, reproducing previously established qualitative conclusions. This demonstrates the practical usefulness of \mpgicrv\ as a quantitative tool for activity indicator selection in multi-GP analyses. 

\section{Conclusions}

We have developed an information criterion for comparing multi-GP models according to their ability to explain the RV component. Exploiting the multivariate Gaussian structure of multi-GPs, we introduce \mpgicrv, which combines the conditional RV likelihood with an effective parameter count. This parameter count incorporates both the explicitly estimated model parameters and the smoothing-based degrees-of-freedom of the GP component, balancing goodness of fit against model complexity in the RV dimension.

We applied \mpgicrv\ to both, synthetic data and several published multi-GP analyses to demonstrate its practical use. In all cases, the criterion quantitatively reproduces expected and previously established qualitative conclusions regarding the relevance of different activity indicators. Informative indicators are favoured through improved regularisation of the RV component, while noise-dominated or unrelated series are penalised.
These results demonstrate that \mpgicrv\ provides a robust, objective, and reproducible framework for multi-GP model comparison.
{  
They also provide a first empirical calibration of the \mpgicrv\ scale. In our tests, models that are nearly indistinguishable by construction typically differ by less than 10 \mpgicrv\ units, while differences $\gtrsim  10$ indicate a clear preference. This supports the AIC-based interpretation adopted in Sect.~\ref{sec:aicformultis}, but the threshold should be treated as indicative rather than universal.
A formal investigation of the \mpgicrv\ threshold as well as its relationship with cross-validation in multi-GP settings is left for future work.}

{  It is also worth to mention that the metrics used to compute \mpgicrv\ may be useful beyond the criterion itself. The conditional likelihood, \lrv, provides a built-in way to evaluate how well different mean functions describe the RV dimension within a multi-GP framework (e.g. to test number of planetary signals). 
Similarly, the smoothing-based degrees of freedom, \ksmooth, provides a direct measure of the effective flexibility of the GP component. Making it useful for comparing different covariance structures, both in one- and multi-GP analyses.
}

Beyond RV applications, the proposed information criterion is general and directly applicable to other multi-GP problems. All equations shown in Sect.~\ref{sec:aicequations} are agnostic of the adopted multi-GP framework. Therefore, this criterion is general and can be applied to any multi-GP problem in which inference targets a specific dimension.

\section*{Acknowledgements}
O.B. acknowledges that has received funding from the European Research Council (ERC) under the European Union’s Horizon 2020 research and innovation programme (Grant agreement No. 865624).
{  O.B. thanks Nathan Hara for his constructive referee report and insightful discussions, which significantly improved the quality and impact of this manuscript.}
O.B. thanks Suzanne Aigrain, Erik Meier, and Thomas Wilson for their valuable comments that help to improve this manuscript. 
This work made use of \texttt{numpy} \citep[][]{numpy}, \texttt{matplotlib} \citep[][]{matplotlib}, \texttt{seaborn} \citep{seaborn} and \texttt{pandas} \citep{pandas} libraries.

\section*{Data Availability}

The codes used in this work are publicly available at \url{https://github.com/oscaribv}. The data analysed in this study are publicly available in the original publications cited throughout the manuscript.

\bibliographystyle{mnras}
\bibliography{refs} 




\appendix

\section{Explicit form of $\mathbf{S}_{\rm rv}$}
\label{ap:hrv}

In this Section we will derive the explicit form of $\mathbf{S}_{\rm rv}$. 
Following equations~\eqref{eq:smoother_full}, \eqref{eq:bigk}, and \eqref{eq:bigkcov} we have
\begin{equation}
    \bm{S}_{\rm big} =
        \begin{pmatrix}
    \bm{k}_{\rm rv,rv} & \bm{k}_{\rm rv,ai}  \\
    \bm{k}_{\rm ai,rv}  & \bm{k}_{\rm ai,ai} \\
    \end{pmatrix}
        \begin{pmatrix}
    \bm{K}_{\rm rv,rv} & \bm{K}_{\rm rv,ai}  \\
    \bm{K}_{\rm ai,rv}  & \bm{K}_{\rm ai,ai} \\
    \end{pmatrix}^{-1}.
    \label{eq:hbigfull}
\end{equation}
Following equation (2.3) in \citet{Lu2002}, and using the conditional covariance of the RVs, $\bm{K}_{\rm rv,rv}^{\prime}$, given in eq.~\eqref{eq:schur}, then eq.~\eqref{eq:hbigfull} can be rewritten as
\begin{equation}
\begin{split}
    \bm{S}_{\rm big} & =
        \begin{pmatrix}
    \bm{k}_{\rm rv,rv} & \bm{k}_{\rm rv,ai}  \\
    \bm{k}_{\rm ai,rv}  & \bm{k}_{\rm ai,ai} \\
    \end{pmatrix} 
    \\
     &   \begin{pmatrix}
    \bm{K}_{\rm rv,rv}^{\prime \, -1} & -\bm{K}_{\rm rv,rv}^{\prime \, -1} \bm{K}_{\rm rv,ai}\bm{K}_{\rm ai,ai}^{-1} \\
    -\bm{K}_{\rm ai,ai}^{-1}\bm{K}_{\rm ai,rv}\bm{K}_{\rm rv,rv}^{\prime \, -1}  & \bm{K}_{\rm ai,ai}^{-1} + \bm{K}_{\rm ai,ai}^{-1} \bm{K}_{\rm ai,rv} \bm{K}_{\rm rv,rv}^{\prime \, -1} \bm{K}_{\rm rv,ai} \bm{K}_{\rm ai,ai}^{-1}\\
    \end{pmatrix}.
\end{split}
\end{equation}

Therefore
\begin{equation}
    \bm{S}_{\rm rv} =
    \bm{k}_{\rm rv,rv} \bm{K}_{\rm rv,rv}^{\prime \, -1} - \bm{k}_{\rm rv,ai}\bm{K}_{\rm ai,ai}^{-1}\bm{K}_{\rm ai,rv}\bm{K}_{\rm rv,rv}^{\prime \, -1}  
    \label{eq:hatrvrvap}
\end{equation}

It is worth noting that in the limit where activity indicators are not connected to the RVs (i.e., $\bm{K}_{\rm rv,ai} = \bm{0}$), then $\bm{S}_{\rm rv}$ reduces to $ \bm{k}_{\rm rv,rv}\bm{K}_{\rm rv,rv}^{-1}$, as expected.

\section{Comparison respect to model 2}
\label{sec:model2comparison}

{ 
Figure~\ref{sec:model2comparison} shows the $\Delta$\mpgicrv\ values from Sect.~\ref{sec:synthetic}, now using Model~2 as the reference. 
Since Model~2 is the true generative model by construction, this provides a direct check of whether \mpgicrv\ identifies the expected model. No alternative model obtains a significantly lower \mpgicrv\ than Model~2, showing that the criterion correctly allow us to detect the best model. 
In addition, models that are nearly indistinguishable by construction typically differ by less than $\sim 5$--$10$ \mpgicrv\ units. This behaviour is consistent with the AIC-based interpretation adopted in Sect.~\ref{sec:aicformultis}, where differences $\gtrsim 10$ are treated as strong evidence for one model over another.
}

\begin{figure*}
\centering
\includegraphics[width=\linewidth]{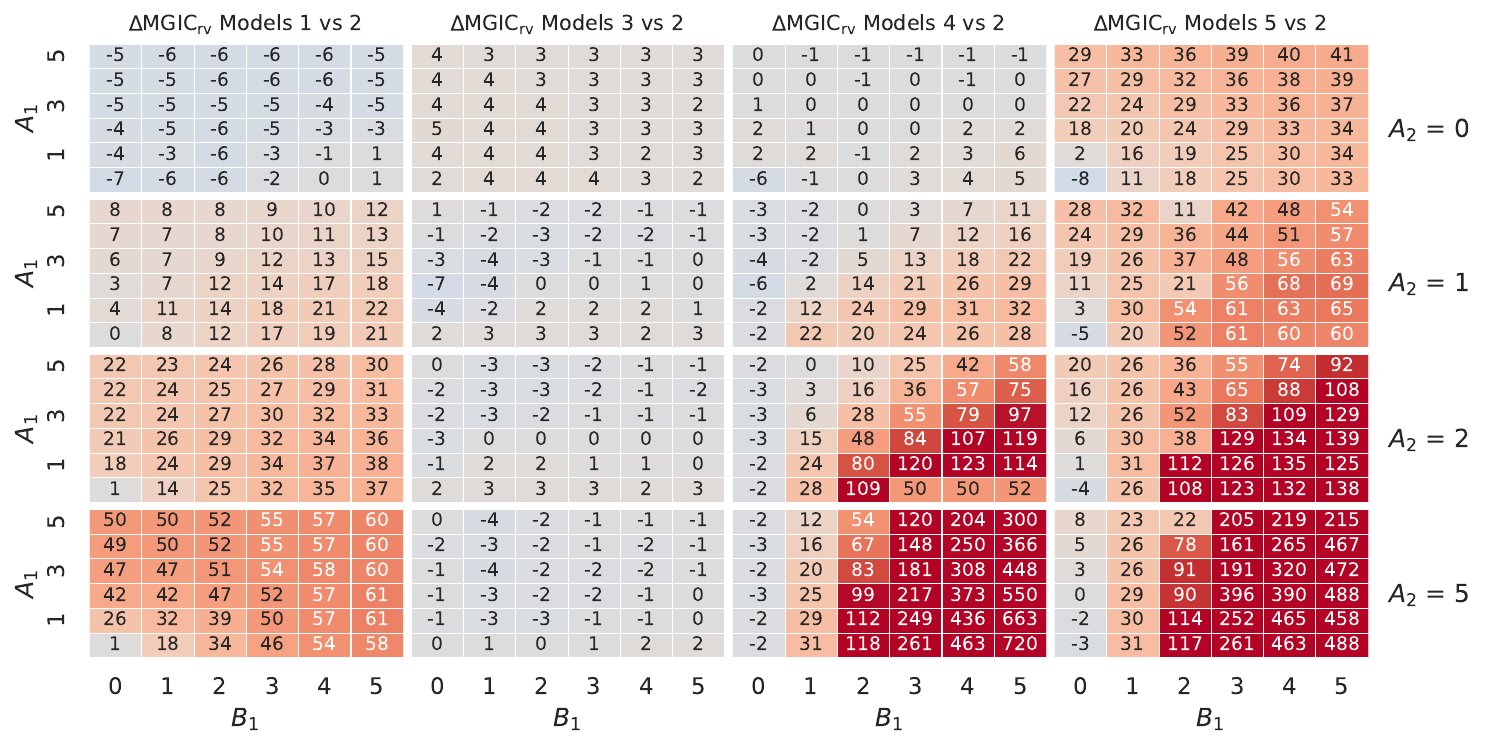}
\caption{Same as Figure~\ref{fig:mgic_all_tests} but taking Model 2 as reference.}
\label{fig:mgic_all_tests-2}
\end{figure*}

\bsp	
\label{lastpage}
\end{document}